\documentclass[aps,showpacs,twocolumn,amsmath,amssymb,amsfonts,eqsecnum]{revtex4-1}
\usepackage{graphicx}
\usepackage{color}
\usepackage{natbib}
\usepackage{epstopdf}
\usepackage{hyperref}

\newcommand{\ket}[1]{|#1\rangle}
\newcommand{\bra}[1]{\langle#1|}
\newcommand{\braket}[2]{\langle#1|#2\rangle}

\begin{document}
\title{Casimir force in the $O(n\to\infty)$ model with free boundary conditions}
\author{Daniel Dantchev$^{1,2}$\thanks{e-mail:
daniel@imbm.bas.bg},Jonathan Bergknoff$^{1}$\thanks{e-mail: jbergk@physics.ucla.edu} and  Joseph Rudnick$^{1}$\thanks{e-mail:
jrudnick@physics.ucla.edu}} \affiliation{ $^1$ Department of Physics and Astronomy, UCLA, Los Angeles,
California 90095-1547, USA,\\$^2$Institute of
Mechanics - BAS, Academic Georgy Bonchev St. building 4,
1113 Sofia, Bulgaria
}
\date{\today}

\begin{abstract}
We present results for the temperature behavior of the Casimir force
for a system with a film geometry with thickness $L$ subject to free boundary conditions and described by the $n\to\infty$ limit of the $O(n)$ model. These results extend over all temperatures, including the critical regime near the bulk critical temperature $T_c$, where the critical fluctuations determine the behavior of the force, and temperatures well below it, where its behavior is dictated by the Goldstone's modes contributions. The temperature behavior  when the absolute temperature,  $T$, is a finite distance below $T_c$, up to a logarithmic-in-$L$ proximity of the bulk critical temperature, is obtained both analytically and numerically; the critical behavior follows from numerics.  
The results resemble---but do not duplicate---the experimental curve behavior for the force obtained for $^4$He films.
\end{abstract}

\maketitle

\section{Introduction}
\label{Section_Introduction}
The Casimir effect remains the object of intense studies,  both in its original formulation due to Casimir \cite{C48}  (see the reviews \cite{GK97,CP2011}), and especially in its thermodynamic manifestation \cite{FG78}---see, e.g, the general reviews \cite{BDT2000,K94} and the reviews devoted to some specific aspects of the critical Casimir force \cite{K99,GD2011,G2009}. The critical Casimir effect has been  directly observed, utilizing light scattering measurements, in the interaction of a colloid spherical particle with a plate \cite{HHGDB2008} both of which are immersed in a binary liquid mixture. In the context of forces that determine the properties of a film of a material in the vicinity of its bulk critical point, the effect has been also studied in $^4$He \cite{GC99},\cite{GSGC2006}, as well as in $^3$He--$^4$He mixtures \cite{GC2002}. Measurements of the Casimir force in thin wetting films of a binary liquid mixture have been performed in \cite{FYP2005} and \cite{RBM2007}. 

Theoretically, the effect has been studies via exact calculations in the two dimensional Ising model \cite{ES94,NN2008,NN2009,AM2010,RZSA2010,DM2013,I2011,WIG2012}, the three dimensional  spherical model \cite{D96,D98,DG2009,DDG2006,CD2004}, with the use of conformal-theoretical methods \cite{A86,BCN86,BE95,ER95,HSED98}, via mean-field type calculations on Ising type \cite{K97,GaD2006,DSD2007,SHD2003,PE92,VMD2011} and $XY$ models \cite{BDR2011}, through renormalization-group studies via $\varepsilon$-expansion of $O(n)$ models \cite{KD92a,KD92b,DGS2006,GD2008,SD2008,DS2011,D2009,D2013}, and via Monte-Carlo calculations \cite{KL96,DK2004,H2007,H2009,H2010,H2011,H2013,VED2013,VGMD2007,VGMD2009,HGS2011,H2012}. In the models envisaged above non-zero critical temperature exists and the thermal fluctuations play the essential role. There are, however, systems in which
the critical point has a quantum origin \cite{SK2011,S2008,S2000,Sa2011} and instead of temperature certain quantum parameters govern the 
quantum fluctuations in the system. In that case one speaks of a quantum critical Casimir effect \cite{CDT2000,BDT2000,PCC2009}.

Given the variety of systems that can exhibit a thermodynamic Casimir effect, the number of measurement techniques that can be applied to its experimental determination and the range of potential applications, it is likely that this state of affairs of large activity in the field of the thermodynamic Casimir effect will persist for some time. 

A recent paper by Diehl \emph{et. al.} \cite{DGHHRS2012} reports on a numerical study of the scaling properties of the thermodynamic Casimir force in thin films  (i.e. dimensions $\infty^2\times L$)  of a Ginzburg-Landau-Wilson (GLW) version of the $O(n)$ model in the limit $n \rightarrow \infty$,  the system being subject to free boundary conditions in the finite direction, focusing particularly on the critical regime immediately below the bulk transition temperature, but also including lower temperatures outside that region. In the current article we extend this study. Our numerical results span the entire range of temperatures, starting from temperatures well below the bulk critical temperature $T_c$, where the Goldstone mode contributions dominate, ranging through the critical regime, where the contributions due to the critical fluctuations of the order parameter dominate, and ending with temperatures far above $T_c$. We confirm  the findings of the authors of Ref. \cite{DGHHRS2012} for the critical regime. In addition, we derive new analytical results for temperatures below $T_c$. By doing so we are able to illuminate the crossover between thermodynamic Casimir forces arising from long-range fluctuations due to Goldstone modes and those arising from critical fluctuations, along the lines of the recent study of Dohm \cite{D2013}. We note that  both types of excitations exist in the low-temperature phase of an $O(n)$-symmetric systems when $n>1$. This phenomenon is thus specific to models with continuous symmetry and does not pertain to Ising type models in which a  discrete symmetry is broken in the ordered state. It is the main reason why the value to which the scaling function of the Casimir force in such models asymptotes below $T_c$ is not zero as in Ising type models, but is rather a nonzero constant \cite{LK91,D96,D98,H2007,VGMD2007,GC99,GC2002,GSGC2006,DGHHRS2012,D2013}. We perform our calculations on a microscopic model---the so-called spherical model \cite{BK52}---that represents the $n\to\infty$ limit of the $O(n)$ models. In contrast with \cite{DGHHRS2012} we do not use the mapping of this model on the GLW  model. Thus the agreement we obtain with \cite{DGHHRS2012} for the critical properties of the model represents a strong manifestation of the validity of the universality hypothesis.  The microscopic formulation of the model is, unlike the GLW approach \cite{note2}, suitable for investigation of the properties of the system at \emph{all} temperatures, particularly at temperatures considerably below that of the bulk transition and thus well outside of the critical regime. 

Because of the continuous symmetry of the model, which is broken at low temperature when $L \rightarrow \infty$, as well as the fact that the boundary conditions correspond to those that are appropriate in the case of $^4$He films, the results of the calculations in \cite{DGHHRS2012}, as well as ours, are qualitatively relevant to the Casimir force measurements on such films described in  \cite{GC99,GSGC2006}. The superfluid transition in $^4$He is, of course, correctly modeled  in terms of  the $XY$, or  $O(2)$, model,  and the results in refs.  \cite{GC99,GSGC2006} have been quite successfully reproduced by Monte Carlo simulations of this model in \cite{H2007} and \cite{VGMD2009}. Nevertheless, the $O(n \rightarrow \infty)$ model merits consideration as a depiction of systems with broken continuous symmetry in the bulk insofar as this model is susceptible to a combination of analytical and numerical approaches, yielding both quantitative and qualitative insights into the behavior of those systems.   
 
We recall that the infinite translational invariant standard spherical model is equivalent to the $n\to\infty$ limit of the corresponding system of $n$-component vectors \cite{S68,KT71,S88,BDT2000,KKPS92,note3}. However, for the spherical model with surfaces or, more generally, without translation-invariant symmetry, this equivalence is preserved only if one imposes spherical constraints in a way which ensures that the mean square value of {\em each} spin of the system is the same \cite{K73}---that is, one averages thermally, but not spatially. Generally such a model is considered analytically intractable. However, as we demonstrate here,  this model can be analytically reduced via exact calculations to a one dimensional model, the properties of which can be then either studied numerically near the critical region, or in an exact analytical manner in the low-temperature regime. The Casimir force within the model  when translational invariance is preserved  have been already studied in \cite{D96,D98} under periodic and in \cite{DG2009} for antiperiodic boundary conditions. There, exact analytical results are derived for the scaling function and the Casimir amplitude for the $d=3$ dimensional film system. 

Results for the quantum version of the spherical model subject to periodic boundary conditions are also available \cite{CDT2000}. Different quantizations of the classical model are possible \cite{BDT2000,O72,V96,VZ92,MVZ97,N95}.  Among them are versions of Bose gas \cite{NJN2013,MVZ97,NP2011,SP85}. Let us also mention the large-$n$ limit of the so-called 2+1 Gross-Neveu model \cite{CT2011}, representative of a broader class of four fermionic models, which lead to mathematics very similar to that of the three dimensional spherical model and to a Casimir amplitude that is exactly equal and opposite to the Casimir amplitude of the three-dimensional spherical model subject to antiperiodic boundary conditions \cite{DG2009}.  The methods utilized here  for the treatment of the spherical model with free boundary conditions may well point the way to progress in the investigation of some of the above-mentioned quantum systems subject to similar boundary conditions; in the references above these models are usually studied in their thermodynamic limit or subject to periodic boundary conditions.

\section{Definition of the model}
\label{Section_Definition}

For an $O(n),n\geq 1$ model of a $d$-dimensional system at a temperature $T$ and geometry $\infty^{d-1}\times L$ the thermodynamic Casimir force per unit area, i.e., the Casimir pressure, is defined by  \cite{E90book,BDT2000}
\begin{eqnarray}
F_{\rm Cas}^{(\tau)}/A & =& F_{\rm Casimir}^{(\tau )}(T,L) \nonumber \\ &=&-\frac{\partial f_{\rm ex}^{(\tau )}(T,L)}{\partial L}%
\text{,}  \label{def}
\end{eqnarray}
where $f_{\rm ex}^{(\tau)}(T,L)$ is the excess free energy per unit area
\begin{equation}
f_{\rm ex}^{(\tau )}(T,L)=f^{(\tau)}(T,L)-Lf_b(T)\text{,}  \label{fexd}
\end{equation}
and the superscript $\tau $ denotes the boundary conditions. Here $f^{(\tau )}(T,L)$ is the full free energy per unit area of such a system subjected to the boundary conditions $\tau $ and $f_b
$ is the bulk free energy density.

Consider a $d$-dimensional cubic lattice,  each lattice site occupied by an $n$-component classical vector spin having ferromagnetic interactions with its nearest neighbors. We single out one dimension, $z$, to be $L$ lattice spacings long. At each of the $L$ sites along the finite dimension, there is a $(d-1)$-dimensional transverse layer containing a total of $A$ spins, where $A$ is large and will later be taken to infinity. Periodic boundary conditions hold within the layers of the system while free boundary conditions are imposed in the $z$ direction by placing a layer of zero length spins on the top and the bottom of the film (i.e., at $z=0$ and $z=L+1$). Since we will consider only such  boundary conditions from here on, the superscript $(\tau)$ will no longer be utilized in the remainder of this article.

The model as described  is not especially amenable to analysis. However, in the $n \rightarrow \infty$ limit it is equivalent\cite{S68,K73} to a form of the spherical model, wherein the vector spins are replaced by real-valued scalar spins and each $(d-1)$-dimensional layer satisfies an individual spherical constraint $\sum s^2=A$, where the summation runs over the spins $s$ belonging to a given layer. We simplify matters further by utilizing the mean spherical model, in which $\langle\sum s^2\rangle=A$, and which yields the same results as the spherical model in the thermodynamic limit, $A\to\infty$.

Our Hamiltonian is therefore
\begin{equation}
\label{MSMHamiltonian}
H=-J\sum_{\langle s,s'\rangle} ss'+J\sum_i\Lambda_i \left(\sum_j s_{i,j}^2-A\right)
\end{equation}
where the first summation is taken over nearest neighbor spins $s$ and $s^{\prime}$, that lie either in the same layer or in adjacent layers. The parameter $J>0$ is the ferromagnetic coupling and $\Lambda_i$ is the ``spherical field'' for layer $i$, i.e. the Lagrange multiplier, which will adjust so as to enforce the mean spherical constraints  $\langle\sum_j s_{i,j}^2\rangle=A$, where the average is taken with respect to the Hamiltonian \eqref{MSMHamiltonian}. The notation $s_{i,j}$ refers to spin $j$ in layer $i$, with $i=1,\ldots,L$ and $j=1,\ldots,A$.

Fourier transforming spins along the layers, periodic boundary conditions being applied, we find
\begin{equation}
H=-JA\sum_i\Lambda_i+\frac{J}{2}\sum_{\mathbf{q}}\mathbf{s}(\mathbf{q})^\dagger\mathcal{H}(\mathbf{q})\mathbf{s}(\mathbf{q})
\end{equation}
where the sum over ${ \mathbf q}$ extends over the first Brillouin zone of layer $i$, and
\begin{equation}
[\mathcal{H}(\mathbf{q})]_{ij}=\mathcal{M}_{ij}-2\delta_{ij}\sum_{k=1}^{d-1}\cos q_k \label{eq:Mintro}
\end{equation}
with 
\begin{equation}
\label{DefinitionOfMMatrix}
\mathcal{M}_{ij}=2\Lambda_i\; \delta_{i,j}-\delta_{|i-j|,1}. 
\end{equation}

After computing the partition function in the standard way, we find the free energy per transverse unit area, in units of $k_B T$, to be
\begin{multline}
\frac{\beta\mathcal{F}}{A}=-R\sum_{i}\Lambda_i+\frac{1}{2}L\ln\left(\frac{R}{2\pi}\right)\\
+\frac{1}{2A}\sum_{\mathbf{q}}\ln\left[\det\left(\mathcal{H}(\mathbf{q})\right)\right],
\end{multline}
where $R=\beta J$. The spherical constraint is enforced in the mean via the Lagrange multipliers $\Lambda_i$. In particular, we must have
\begin{equation}
\label{SphericalConstraintArbitraryDimension}
0=\frac{\beta}{A}\frac{\partial\mathcal{F}}{\partial\Lambda_i}=-R+\frac{1}{A}\sum_{\mathbf{q}}\left[\mathcal{H}(\mathbf{q})\right]^{-1}_{ii}
\end{equation}
for each $i=1,\ldots,L$.

\section{Results on the model in $d=3$}
\label{Section_Results}

\begin{figure*}[t]
\begin{center}
\includegraphics[width=6.5in]{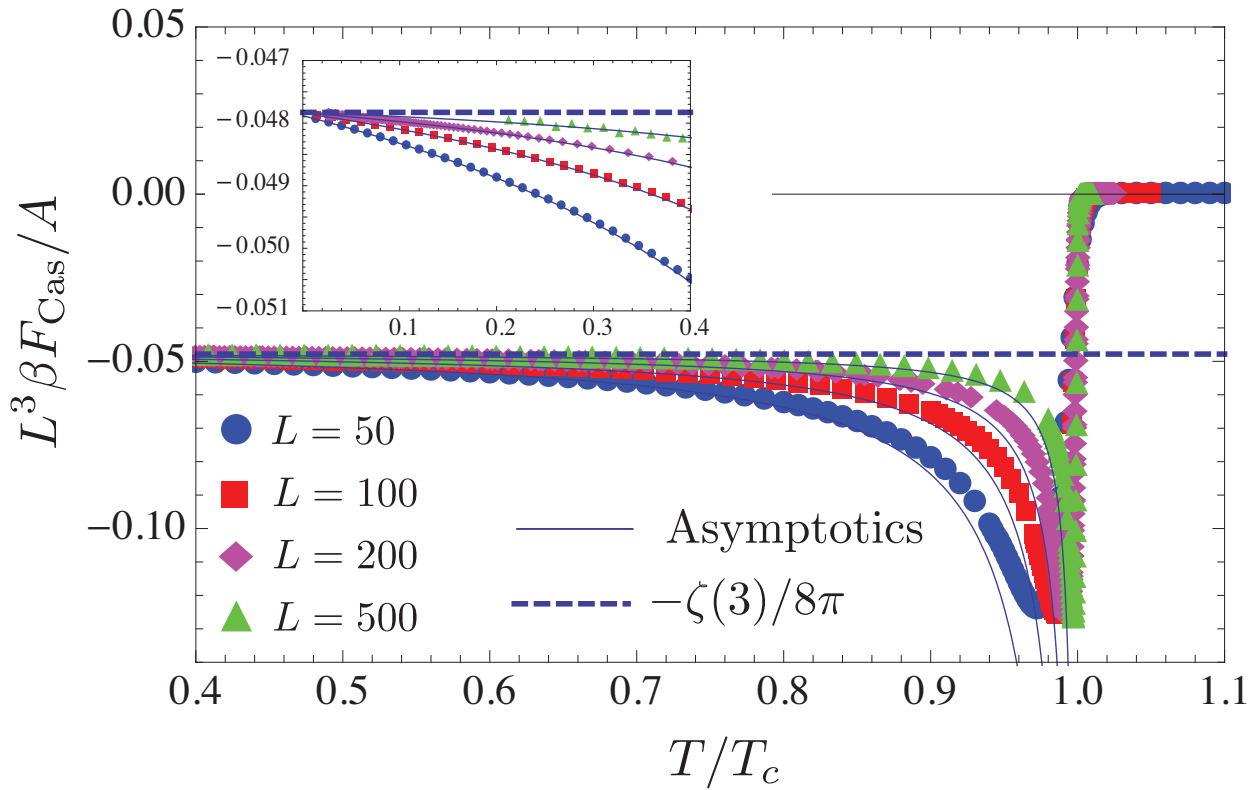}
\end{center}
\caption{(Color online) The scaled Casimir force (symbols) as compared to the closed form low temperature asymptotic results (solid curves) developed in section \ref{Section_Asymptotics} for $L=50, 100, 200$ and 500.  The asymptotic results turn out to be accurate for moderately low absolute temperatures, corresponding to  $T \leq 0.8 \,T_c$. Note that larger $L$ better the approximation given by the asymptotic curves, as should be expected, since $4\pi(R-R_c)L\gg \ln L$ is the variable that governs the behavior observed. As indicated in the inset, which tracks the scaled Casimir force down to absolute zero, the asymptotic forms, given by Eqs. \eqref{d1} and \eqref{d2}, are quite accurate at lower temperatures for any $L\gg 1$.}
\label{Figure_AsymptoticComparisonPlot}
\end{figure*}

\begin{figure}[b]
\begin{center}
\includegraphics[width=\columnwidth]{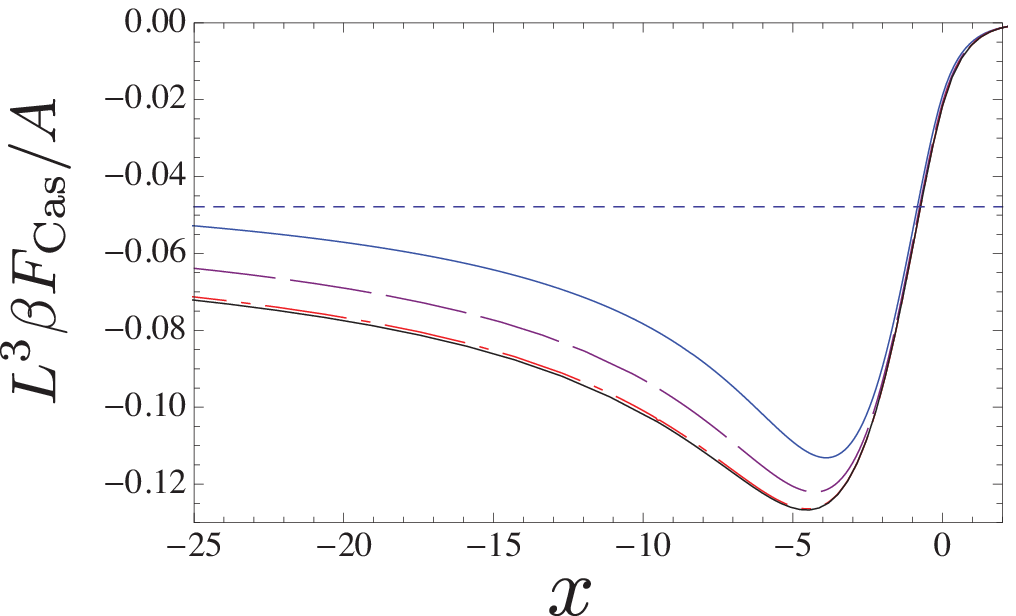}
\end{center}
\caption{(Color online) The scaled Casimir force $L^3\beta F_{\textrm{Cas}}/A$ as a function of scaling variable $x=(L/\xi_0^+)t$, with reduced temperature $t=(T-T_c)/T_c$ and bulk correlation length amplitude $\xi_0^+$, for $L=10$ (top, solid blue), $L=30$ (second from top, dashed purple), $L=200$ (second from bottom, dot-dashed red), $L=500$ (bottom, solid black). The zero-temperature value of $-\zeta(3)/8\pi$ is indicated as a horizontal dashed line. The $L=200$ and $L=500$ curves lie on top of eachother in the critical region $x=O(1)$, and both are close approximations to the scaling function $\vartheta(x)$.}
\label{Figure_Casimir_Several_L}
\end{figure}

We now focus on the case of three dimensions. As a prelude to this discussion, we display in Fig. \ref{Figure_AsymptoticComparisonPlot} our results for the Casimir force for an extended temperature range, from well below the bulk transition temperature, $T_c$ to just above it. The horizontal axis is $T/T_c$, and the vertical axis is the scaled Casimir force per unit area, $L^3\beta F_{\rm Cas}/A$. We choose the scale factor $L^3$ because in systems with broken continuous symmetry the Casimir force scales as $L^{-3}$ both below and in the vicinity of $T_c$ and decays exponentially above that temperature \cite{note4}.
The behavior of the force near $T_c$ is presented in Fig. \ref{Figure_Casimir_Several_L}. We observe that the results plotted in Figs. \ref{Figure_AsymptoticComparisonPlot}, \ref{Figure_Casimir_Several_L}  and \ref{Figure_Casimir_scaled} agree with the expected  behavior of the Casimir force in systems with broken continuous symmetry \cite{K94,BDT2000}. Specifically, in this system we find that the Casimir force scales as $L^{-3}$ both well below and near $T_c$; the scaling function of the force tends to a nonzero constant for $x\to-\infty$; and the force is negative, i.e., a force of attraction for all temperatures, as one expects when the boundary conditions are the same at both bounding layers.  

The calculations leading to the results displayed in the two figures are as follows. We start with the one dimensional operator 
\begin{equation}
\label{HOuterProductExpansion}
\mathcal{H}(\mathbf{q})^{-1}=\sum_{l=1}^L\frac{\ket{\psi^{(l)}}\bra{\psi^{(l)}}}{a_l-2\cos q_x-2\cos q_y},
\end{equation}
where $\{a_l,\ket{\psi^{(l)}}\}$ are the eigenvalues and normalized eigenvectors, respectively, of the matrix $\mathcal{M}$ defined in Eqn. (\ref{DefinitionOfMMatrix}). Note that they both $\{a_l\}$ and $\{\ket{\psi^{(l)}}\}$  depend on the spherical fields $\Lambda_i$, $i=1,\cdots,L$. The general properties of $\{a_l\}$ and $\{\ket{\psi^{(l)}}\}$, with $l=1,\cdots,L$ are given in Appendix \ref{sec:matrix_M}. Here we note that all eigenvalues $a_l$ are real, non-degenerate and, if $a_1$ is the single ground-state value one has $a_1>0$ and that the  corresponding eigenvector can be chosen to have positive components, i.e., $\psi^{(1)}_i > 0$ for all $i=1,\cdots,L$. 

Given $\{a_l\}$ and $\ket{\psi^{(l)}}$, $l=1,\cdots,L$  with Eq. (\ref{SphericalConstraintArbitraryDimension}) satisfied, we are in a position to determine all the thermodynamic properties of this system.  In the transverse thermodynamic limit, $A\to\infty$, the sum over Brillouin zone is reproduced by an integral, and Eqs. (\ref{SphericalConstraintArbitraryDimension}) and (\ref{HOuterProductExpansion}) lead to the so-called spherical constraints
\begin{eqnarray}
\label{SphericalConstraintThreeDimensions}
\beta J & \equiv & R \nonumber \\ &=& \sum_{l=1}^L\left[\psi^{(l)}_i\right]^2 g(a_l),
\end{eqnarray}
for each $i=1,\ldots,L$, where $\psi^{(l)}_i$ is the $i$-th component of the eigenvector  $\ket{\psi^{(l)}}$, 
\begin{equation}
\label{gf_def}
g(x)=\frac{1}{2\pi}\frac{4}{x}K\left(\frac{4}{x}\right),
\end{equation}
and $K(k)$ is the complete elliptic integral of first kind with modulus $k$. Using the completeness of the eigenvectors $\ket{\psi^{(l)}}$, $l=1,\cdots,L$ and performing the sum of Eqs. \eqref{SphericalConstraintThreeDimensions} with respect to $i$, we arrive at
\begin{equation}
\label{SummingSphericalConstraintEquationsshort}
R=\dfrac{1}{L}\sum_{l=1}^L g\left(a_l\right).
\end{equation}

The free energy can then be written in closed form as
\begin{multline}
\label{FiniteSizeFreeEnergy}
\frac{\beta\mathcal{F}}{A}=\frac{1}{2}L\ln\left(\frac{R}{2\pi}\right)+\frac{1}{2}\sum_{l=1}^L\Bigg[\ln a_l-2R\Lambda_l\\
-\frac{2}{a_l^2}\cdot{}_4F_3\left(1,1,\frac{3}{2},\frac{3}{2};\,2,2,2;\,\frac{16}{a_l^2}\right)\Bigg],
\end{multline}
where ${}_4F_3$ is a generalized hypergeometric function \cite{AS}. The bulk free energy per length, $f_b$, is straightforwardly calculated using known methods---see Appendix \ref{sec:bulk_model}.

Then,  as in \cite{DGHHRS2012}, we construct the Casimir pressure
\begin{multline}
\frac{\beta F_{\textrm{Cas}}}{A}=-\frac{\partial}{\partial L}\left(\frac{\beta\mathcal{F}}{A}-L\beta f_b\right)\\
\approx \beta f_b-\frac{1}{2}\left[\frac{\beta\mathcal{F}(L+1)}{A}-\frac{\beta\mathcal{F}(L-1)}{A}\right], \label{ourexpression}
\end{multline}
where $f_b$ is the bulk free energy density---see Eq.  \eqref{BulkSphericalFreeEnergyDensity}. 
Implementing the analysis described above, with the eigenvalues $\{a_l\}$ and the eigenvectors $\ket{\psi^{(l)}}$, $l=1,\cdots,L$  determined from the matrix $\mathcal{M}$ where $\Lambda_i$, $i=1,\cdots,L$  are determined to satisfy Eq. (\ref{SphericalConstraintArbitraryDimension}) with the use of the numerical methods described in Appendix \ref{Numerics}, we find the Casimir force curves shown in Figs. \ref{Figure_AsymptoticComparisonPlot} and  \ref{Figure_Casimir_Several_L}. 

\subsection{Behavior of the critical Casimir force}
\label{critical_casimir_force}

\begin{figure*}[htbp]
 \begin{center}
 \includegraphics[width=6.5in]{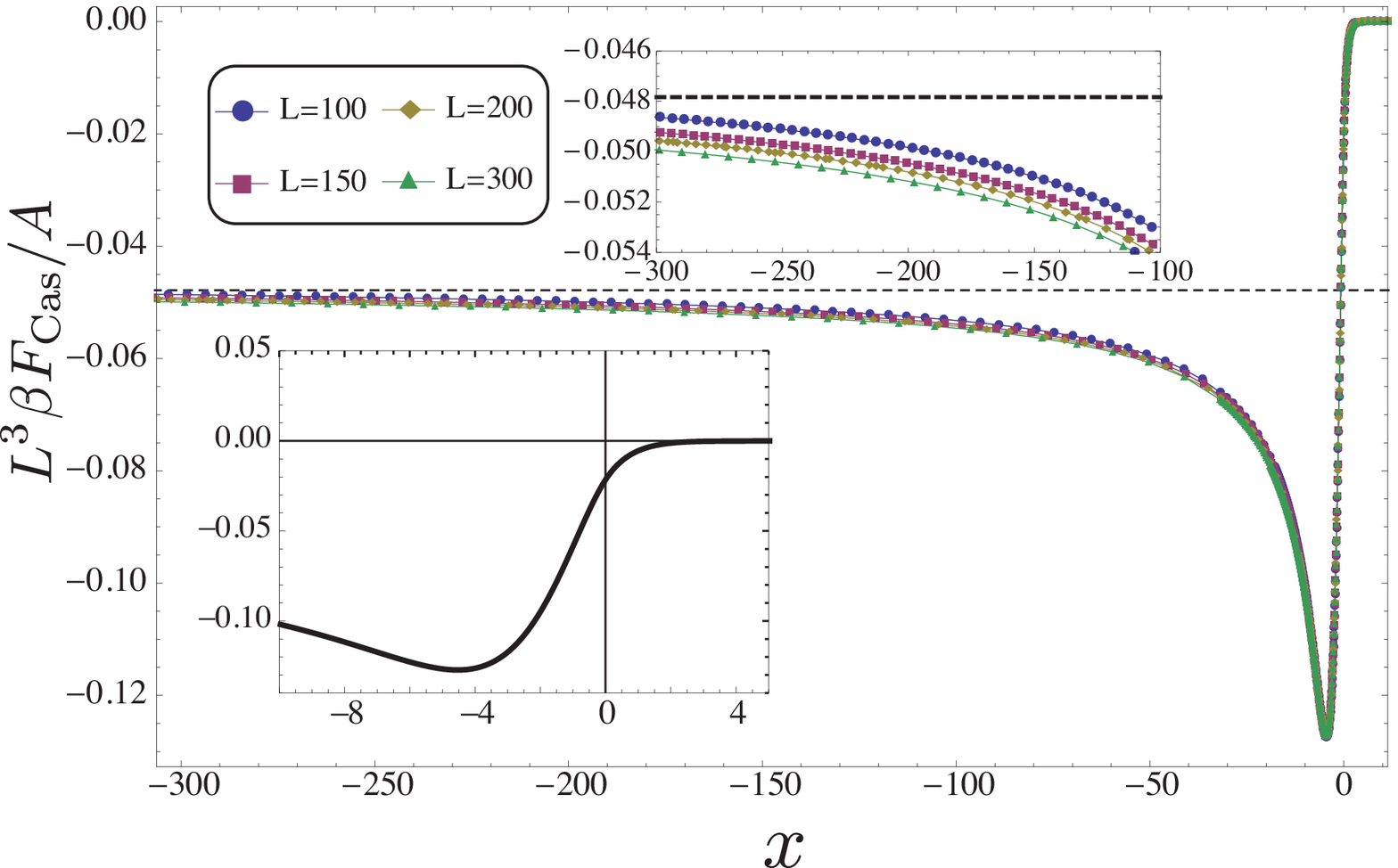}
 \end{center}
 \caption{(Color online) The scaled Casimir force $L^3\beta F_{\textrm{Cas}}/A$ as a function of scaling variable $x=(L/\xi_0^+)t$ in the temperature region close to and well below the critical temperature $T_c$ allowing for only linear-in-$L$ corrections to scaling (see the discussion in the last two paragraphs of Sec. \ref{critical_casimir_force}).  Data for $L=100,150,200$ and $L=300$ are presented. The notations are the same as in Fig. \ref{Figure_Casimir_Several_L}. The lower inset shows a blow-up of the region close to $T_c$ and demonstrates the excellent scaling there that can be achieved in this way, in that all curves are indistinguishable. The upper inset shows blow-up of the region $x\in(-300,-100)$ and depicts the spreading of the scaling curves in the regime well below $T_c$ due to the existence of a logarithmic-in-$L$ term there; see Eq. \eqref{d2}. The zero-temperature value of $-\zeta(3)/8\pi$ is indicated as a horizontal dashed line.}
 \label{Figure_Casimir_scaled}
 \end{figure*}

 Figures \ref{Figure_Casimir_Several_L} and \ref{Figure_Casimir_scaled} displays the scaled critical Casimir force.  
The scaling variable is $x= (L/\xi_{0}^+) t$, where $t=(T-T_c)/T_c$ is the reduced temperature and $\xi_b(t\to 0^+)=\xi_{0}^+ t^{-\nu}$ is the bulk correlation length. Here, $\nu=1$ is the corresponding critical exponent in the bulk spherical model in $d=3$, and $\xi_0^+=(4\pi R_c)^{-1}$ the non-universal amplitude as determined from earlier results \cite{D96}. Here $R_c$ is the bulk critical coupling
 \begin{equation}
 \label{bulk_constraint}
 R_c=\dfrac{1}{2}\dfrac{1}{(2\pi)^3}\int_{-\pi}^\pi \frac{d^3q}{3-\cos q_x-\cos q_y-\cos q_z}.
 \end{equation}
 In \cite{JZ2001} it has been shown that 
 \begin{equation}
 \label{JCart}
 R_c=\frac{\left(\sqrt{3}-1\right) \Gamma \left(1/24\right)^2 \Gamma
    \left(11/24\right)^2}{192 \pi ^3}\backsimeq 0.252731.
 \end{equation}
 According to \cite{W39}, the above is also equivalent to
 \begin{eqnarray}
 \label{W39Rc}
 R_c &=& \frac{4}{\pi ^2} \left(18+12 \sqrt{2}-10 \sqrt{3}-7 \sqrt{6}\right) \\
 && \times \;
    K\left[\left(2-\sqrt{3}\right)
    \left(\sqrt{3}-\sqrt{2}\right)\right]^2. \nonumber
 \end{eqnarray}

  As $L\to\infty$, corrections to scaling become negligible and we are left with the Casimir scaling function $\vartheta(x) = L^3 \beta F_{\rm Cas}(x)/A$ for this system under free boundary conditions. The curves for $L=200$ and $L=500$ are substantially the same; the solid black $L=500$ curve is, in fact, an excellent approximation to $\vartheta(x)$.

From our numerical results for a set of $L$ values, $L\in \{10,20,30,40,50,60,70,80,90, 100,150,200, 250,300,400,\\500, 750, 1000,1250,1500, 1750,2000,3000\}$, we find a Casimir amplitude of
\begin{equation}
\Delta=\frac{1}{2}\vartheta(0)\approx -0.010773(7).
\end{equation}
The reported value represents a conservative estimate of the constant, arrived at by fitting the data with corrections to scaling that are either logarithmic in $L$, or purely linear in $L$. To be specific, we fitted the data with corrections to the leading behavior of the form $\sum_i^{n_{\rm max}} (c_i +d_i\ln L)L^{-i}$, where we have taken $n_{\rm max}=5$, and the coefficients  $d_i$ have been either determined by a least squares procedure or set equal to zero. The  values for the Casimir amplitude by the two approaches are in close agreement, and the value reported above is consistent with what we find by either of the two approaches. When the coefficients $d_i$ are allowed to adjust, the leading coefficient, $d_1$ turns out to be quite small. Based on this, we feel that we can neither confirm noor refute, the existence of logarithmic corrections in the behavior of $\Delta$.

The extremum of the scaling function and its position is determined for a set of values $L\in \{10,20,30,40,50,60,70,80,90, 100,150,200, 250,300,400,\\500\}$. We follow the same procedure as  was utilized to determine  the Casimir amplitude $\Delta$. The result we obtain for the minimum value of the force is 
\begin{equation}
\vartheta_{\textrm{min}}\approx-0.1270(2).
\end{equation}
 The amplitude $\Delta$ was first evaluated in \cite{CHG2009} where the value   $\Delta=-0.012(3)$ was reported. The location of the Casimir force extremum is at $x=x_{\rm min}=-4.53$ for $L=500$. The conservative estimate obtained from the set of all $L$ values is $x_{\rm min}=-4.54(1)$. 

The quantities $\Delta,\vartheta_{\textrm{min}}$ and $x_{\rm min}$ are expected to be universal and, indeed, they agree to great precision with the values given by Diehl, \emph{et. al.} \cite{DGHHRS2012} for their closely related model. The accuracy of our results are limited by our approximation of the derivative in (\ref{def}) by a difference (see Eq. (\ref{ourexpression})). Agreement with the measurements in \cite{GC99,GSGC2006} for $^4$He films is less satisfactory, as it should be, given the widely-acknowledged difference between the $O(2)$ and $O( \infty)$ models; experimental results on $^4$He films are consistent with $x_{\rm min}=-5.7(5)$ and $\vartheta_{\textrm{min}}=-1.30(3)$ \cite{GSGC2006}.

Fig. \ref{Figure_Casimir_scaled} depicts the scaled Casimir force $L^3\beta F_{\textrm{Cas}}/A$ as a function of scaling variable $x=(L/\xi_0^+)t$ in the temperature region close to and well below the critical temperature $T_c$ allowing for only linear-in-$L$ corrections to scaling. These corrections amount to replacing the film thickness $L$ by an effective thickness $L_{\rm eff}=L+\delta L$, where $\delta L$ does not depend on $L$ and $T$.  Since $\nu=1$ for the three-dimensional (bulk) spherical model, the last replacement of $L$ with $L_{\rm eff}$ means taking into account the linear in $L$ corrections to scaling. This procedure is consistent with the essential ambiguity in the lateral size of a lattice system with free boundary conditions, in that it is not clear what portion of the space outside the boundary layers of the system ought to be attributed to the system itself. It is reasonable to expect the uncertainty to be of the order of a lattice spacing, which here is equal to 1. In Fig. \ref{Figure_Casimir_scaled} data for $L=100, 150, 200$ and $L=300$ are presented. It turns out that $\delta L=0.45$ leads to a near perfect overlap of the curves near $T_c$. 

The notations in the figure are the same as in Fig. \ref{Figure_Casimir_Several_L}. The lower inset is a blow-up of the region close to $T_c$ and demonstrates the agreement with scaling that has been achieved with the use of $L_{\rm eff}$, in that all curves are indistinguishable. The upper inset shows blow-up of the region $x\in(-300,-100)$ and depicts the spreading of the scaling curves in the regime well below $T_c$ due departure from finite size scaling. This violation of the scaling hypothesis can be traced to the existence of a logarithmic-in-$L$ term there. Why there is such spreading, why there logarithmic-in-L corrections exist and what are their amplitudes will be derived in Section \ref{Section_Asymptotics}. The zero-temperature value of $-\zeta(3)/8\pi$ is indicated as a horizontal dashed line. This asymptotic value and the leading $L$-dependent corrections to it are also derived in the next section.

\subsection{The Casimir force in the Goldstone mode dominated regime}
\label{Section_Asymptotics}

In the regime in which Goldstone modes dominate, i.e. for $T \ll T_c$ (or, equivalently, $R\gg R_c$) the system explored here can be studied in closed form. In that regime, the left hand side of Eqn. (\ref{SphericalConstraintThreeDimensions}) becomes large, forcing the lowest eigenvalue, $a_1$, to approach $4$ and dominate the right hand side, since then $K(x)$ grows logarithmically according to \cite{AS}: $K(x)\simeq \ln{[4/\sqrt{1-x^2}]}$. This  causes the summation over $l$ to be dominated by the $l=1$ contribution and,  as the right hand side of  (\ref{SphericalConstraintThreeDimensions}) must be independent of  the site index, we have $\left.\psi^{(1)}_i\right.^2$ equal to a constant, independent of $i$. Numerically, one can check that $\psi^{(1)}$ approaches a constant vector. From the fact that $\psi^{(1)}$  is the eigenvector of $\mathcal{M}$ with eigenvalue $4$, we find 
\begin{eqnarray}
\mathbf{\Lambda}&=&\mathbf{\Lambda}^* \nonumber \\ &=&
(5,6,6,\ldots,6,6,5)/2 \label{lamst}
\end{eqnarray}
Since the matrix $\mathcal{M}^*\equiv\mathcal{M}(\mathbf{\Lambda}^*)$ can be directly expressed in terms of the discrete Laplacian under Neumann-Neumann boundary conditions for the eigenvalues $\lambda_l$ and normalized eigenvectors $\phi^{(l)}$ of $\mathcal{M}^*$ one immediately has, see, e.g., Ref. \cite{BDT2000}
\begin{equation}
\lambda_l=4+4\sin^2\left(\pi(l-1)/2L\right) \label{eq:aslams}
\end{equation}
and 
\begin{equation}
\phi^{(l)}_i = \left\{\begin{array}{ll} 1/\sqrt{L}, & i=1 \\ 
\cos \left[ \pi(l-1)(i-1/2)/L \right]/\sqrt{L/2},  & i = 2, \ldots , L  \end{array} \right. \label{eq:pert4}
\end{equation}
where   $\phi^{(l)}$ are orthonormal and form a complete system.  We then expand about $\mathbf{\Lambda}=\mathbf{\Lambda}^*$ applying perturbation theoretical methods in the small variable $1/L(R-R_c)$.  Using the constraints, Eqs. (\ref{SphericalConstraintThreeDimensions}) and \eqref{SummingSphericalConstraintEquationsshort}, we find
\begin{equation}
\label{a1}
a_1\simeq 4+32e^{-4\pi L(R-R_c)}, 
\end{equation}
nonperturbatively, i.e., inexpressible as a power series in $1/L(R-R_c)$, and
\begin{eqnarray}
\label{al}
a_l&=&\lambda_l+\frac{1}{2\pi L(R-R_c)}(\lambda_{2l-1}-4)  \\
&&\times\left[\frac{K(4/\lambda_l)}{\lambda_l}-\frac{K(4/\lambda_{L+2-l})}{\lambda_{L+2-l}}\right]+O\left([(R-R_c)L]^{-2}\right) \nonumber
\end{eqnarray}
perturbatively, for $l\ge 2$, keeping only the first order correction with respect to the variable $1/L(R-R_c)$. The details of the derivation of Eqs. \eqref{a1} and \eqref{al} are presented in Appendix \ref{sec:low_T}. Note that Eq. \eqref{a1} demonstrates that $a_1\to 4$ when $4\pi L(R-R_c)\gg 1$. This also determines the range of the validity of Eq. \eqref{al}. The results presented in Eqs. \eqref{a1} and \eqref{al} can be further refined as shown in in Appendix \ref{sec:low_T}---see Eq. \eqref{rolLimproved}---by replacing $R_c$ in them with $\rho_L$, where 
\begin{equation}
\label{rolLimprovedMT}
\rho_L=R_c-\frac{1}{4\pi L}\left(\frac{K(1/2)+7\ln 2}{2}+\ln L\right)+O\left(\frac{1}{L^2}\right).
\end{equation}
The expressions \eqref{a1} and  \eqref{al} for $a_1$ and $a_l$, $l=2,\cdots,L$ then become
\begin{equation}
\label{a1_final}
a_1=4+\frac{1}{L} 2 \sqrt{2} e^{-K\left(1/2\right)/2} e^{-4\pi 
   \left(R-R_c\right)L},
\end{equation}
and 
\begin{equation}
\label{al_final}
a_l=\lambda_l+\frac{\sin^2[\pi(l-1)/L]}{L(R-\rho_L)}
\left[g(\lambda_l)-g(L+2-l)\right].
\end{equation}
As shown in Appendix \ref{sec:low_T}, the above equations are valid for
\begin{equation}
\label{g1g2}
4\pi (R-R_c)\gg \ln L/L,
\end{equation}
This means that our ``low temperature'' calculations are accurate to a distance below the critical point going as $\ln L/L$, which is well outside the region in which critical point scaling hods ($(R-R_c) \sim 1/L$), but nevertheless quite close on an absolute temperature scale. In this latter regime, $a_1$ approaches 4 as a function of $L$ faster than $L^{-2}$. Obviously, if $T$ is at a fixed, $L$-independent distance below $T_c$ then $a_1\to 4^+$ exponentially rapidly in $L$.

In order to compute the Casimir force, we must determine the effects of the perturbed eigenvalues on the free energy, Eq. (\ref{FiniteSizeFreeEnergy}). Given the above discussion, when $4\pi L(R-R_c)\to \infty$ our system will behave like the corresponding Gaussian model under Neumann-Neumann boundary conditions at its critical point. This leads to the well known result, see, e.g., Ref. \cite{KD92a}, $\beta F_{\textrm{Cas}}=-\zeta(3)/(8\pi)L^{-3}$. It is interesting to note that one has the same result also for the Gaussian model under Dirichlet-Dirichlet boundary conditions \cite{KD92a}. The analysis of the case in which $4\pi L(R-R_c)$ is large but finite is much more involved. The details are contained in Appendix \ref{sec:low_T}. The result is that the force $\beta F_{\rm Cas}$ can be represented as a sum of a leading order, temperature-independent term $\beta F_{\rm Cas}^{(0)}(L)$, plus a term that reflects the leading temperature-dependent contributions $\beta F_{\rm Cas}^{(1)}(T,L)$.  One can derive an {\it exact} expression for  $\beta F_{\rm Cas}^{(0)}(L)$. The result is 
\begin{eqnarray}
\label{F0}
\lefteqn{\beta F_{\rm Cas}^{(0)}(L)=}\\
&& -\frac{1}{(2\pi)^2}\int_{-\pi}^\pi\int_{-\pi}^\pi dq_x dq_y \frac{v(q_x,q_y)}{\exp[2Lv(q_x,q_y)]-1}, \nonumber
\end{eqnarray}
where
\begin{equation}
\label{v}
v(q_x,q_y)=\cosh^{-1}\left[3-\cos q_x-\cos q_y\right].
\end{equation}
Obviously, $\beta F_{\rm Cas}^{(0)}(L)<0$. Expanding  $\beta F_{\rm Cas}^{(0)}(L)$ in powers of $1/L$, we find
\begin{multline}
\label{d1}
\frac{\beta F_{\rm Cas}^{(0)}(L)}{A}=-\frac{1}{8\pi L^3}\Bigg[\zeta(3)+\frac{7}{8}\zeta(5)L^{-2}
+O\left(L^{-4}\right)\Bigg],
\end{multline}
while for $\beta F_{\rm Cas}^{(1)}(T,L)$ one has 
\begin{equation}
\label{d2}
\frac{\beta F_{\rm Cas}^{(1)}(T,L)}{A}=-\frac{1}{4(R-R_c)L^4}\Big[a+b\ln L+O\left(L^{-2}\right)\Big],
\end{equation}
with
\begin{multline}
a=\frac{\zeta'(-2)}{4}\left(2+3K\left(\frac{1}{2}\right)-21\ln 2+6\ln(2\pi)\right)\\
-\frac{3\zeta''(-2)}{4} \simeq 0.0224639  \label{eq:aform}
\end{multline}
and
\begin{equation}
b=-\frac{3\zeta'(-2)}{2}\simeq 0.0456727 , \label{eq:bform}
\end{equation}
where $\zeta$ is the Riemann $\zeta$-function. Note that $1/(R-R_c)\sim T$ for $T\approx 0$ and, thus, Eq. \eqref{d2} can be safely used even at very low temperatures. In addition, Eqs. \eqref{d2}-\eqref{eq:bform} imply that  $\beta F_{\rm Cas}^{(1)}(T,L)<0$.

The fact that the quantity $\beta F_{\rm Cas}/A$ is negative at $T=0$ and that it decreases with increasing $T$, eventually approaching zero at temperatures just above the bulk critical temperature, tells us that there must be at least one minimum in that quantity in the temperature range $0<T<T_c$. We find precisely one such minimum. The above implies that the Casimir force in the  model considered here is nonmonotonic as a function of $T$, as opposed to its behavior under periodic boundary conditions \cite{D96,D98}---in which case the Casimir force has beeb analytically proven to be monotonically increasing; in the case of antiperiodic boundary conditions \cite{DG2009} the force is a monotonically decreasing function of $T$. We can thus associate the non-monotonicity with the existence of physical bounding surfaces in the system subject to free boundary conditions. This property is also observed in the $XY$ model \cite{H2007,VGMD2007} subject to Dirichlet boundary conditions and may well persist in any $O(n)$, $n \ge 2$ model under boundary conditions enforcing the existence of surfaces in the geometry of the system.

The comparison between numerical and asymptotic results is shown in Fig. \ref{Figure_AsymptoticComparisonPlot}. We observe good agreement between them for $T\le 0.8 T_c$ for the $L$ values considered there---the larger $L$ the better the agreement, as should be expected, since $(R-R_c)L$ is the variable that governs the behavior observed for large values of that variable. 

Interestingly, we find that our analytical and numerical results at low temperature are inconsistent with those reported in \cite{DGHHRS2012}. In particular, our expressions (\ref{d1})--(\ref{eq:bform}) are inconsistent with the low temperature behavior plotted in their Fig. 1, especially in the sense that our results do not collapse into a scaling form expressible entirely in terms of the combination $x=(L/\xi_0^+)t$. Furthermore, we find that the expression utilized by them (see the caption of their Fig. 1) does not reproduce our low temperature results. This discrepancy may arise from inconsistencies between their low temperature approach and ours. Of course, low temperature behavior may well be model-dependent. Nevertheless, this regime deserves further exploration. 

Finally, our results delineate the regions in which critical fluctuations and Goldstone modes dominate the behavior of the Casimir force: Eq. \eqref{g1g2}, which can be also rewritten as 
$-x R/R_c\gg \ln L$, yields the condition on $T$ and $L$ in which the Goldstone contributions dominate, while $x=O(1)$ is the finite-size scaling critical region in which the critical fluctuations dominate. Of course, when  $\ln L \gg -x R/R_c \gg 1$, which defines a region near $T_c$, both the critical fluctuations and Goldstone type excitations mix so that neither of them dominate. The  validity of these results beyond the specific model investigated here remains to be determined.
 
\section{Conclusions and discussion}

We have found that the venerable spherical model  \cite{BK52,LW52}, which has proven so useful in the reproduction and elucidation of thermodynamic behavior in a number of interesting systems (for a review see Refs. \cite{J72,KKPS92,F2005,BDT2000}) provides insight into the critical Casimir force in a system having a broken continuous symmetry in its ordered state. Most of the studies of this model have been performed for systems in which translation-invariant symmetry is present, in which case the model is equivalent to the $n\to\infty$ limit of the corresponding $n$-component vector models \cite{S68,KT71}. However, for the spherical model with surfaces  this equivalence is preserved only if one imposes spherical constraints in a way which ensures that the mean square value of {\em each} spin of the system is the same \cite{K73}, a version of the model that was viewed for some time as analytically intractable \cite{BJSW74,BM77}. In the current study we were able to implement the properly formulated spherical model to extract analytical results at temperatures below of the critical region---see Eq. \eqref{g1g2}, and in addition, provide indications that it may be possible to derive exact results at the critical point. Thus, our approach, which confirms the results reported by Diehl \emph{et. al.} \cite{DGHHRS2012} near $T_c$ and extends and partially corrects them in the region below $T_c$ (see Sec. \ref{Section_Asymptotics}) provides insight into the connection between a Casimir force in a film when it is driven by critical fluctuations in the immediate vicinity of the bulk phase transition and a Casimir force that reflects the influence of Goldstone modes at lower temperatures.  

While the Casimir forces obtained with the use of this model, as displayed in Figs. \ref{Figure_AsymptoticComparisonPlot}, \ref{Figure_Casimir_Several_L},  and \ref{Figure_Casimir_scaled} differ in detail from the data for $^4$He obtained in \cite{GC99,GSGC2006}, the overall features---particularly the pronounced minimum in the Casimir force below the critical point and the approach to a non-zero limiting value at low temperatures---are strikingly similar. As noted above, the low-temperature behavior---see Eqs. (\ref{d1}) and (\ref{d2})---reflects the Goldstone mode contributions to the Casimir force, the leading behavior of which is given by $-\zeta(3)/(8\pi)$ \cite{LK91}. We note that the formulation of the model explored here fails to capture hydrodynamic surface wave fluctuations, which play a role in the low temperature Casimir force of a film of superfluid liquid \cite{ZRK2004}. 


Equations \eqref{rolLimprovedMT}--\eqref{al_final} suggest the existence of an additive logarithmic shift to the scaling variable in the region near $T_c$, for which  $\ln L \gg -x R/R_c \gg 1$, where both the critical fluctuations and Goldstone type excitations mix so that neither of them dominate---see Eq. \eqref{g1g2}. In that regime one is outside the finite-size critical regime, since $|x|\gg 1$, but still not in the Goldstone dominated regime.  We do not have analytical results to support---or to refute---the proposition that this behavior persists into the critical regime or at $T_c$. Extending the method in Appendix \ref{sec:low_T},  one can contemplate developing a perturbation theory in which $a_1\to 4^+$ but the constraint $g(a_1)\gg g(\lambda_2)$ is abandoned. Numerical results indicate that at $T=T_c$ one has $a_1-4\propto L^{-2}$ \cite{pert}. Finally, we have shown that when $\ln L\ll -x R/R_c$, see Eq. \eqref{g1g2}, there are $\ln L$ corrections to the behavior of the Casimir force (see Eq. \eqref{d2}); that is, the leading temperature dependence cannot be expressed entirely in terms of the scaling variable $x$. It should be possible to utilize Monte Carlo simulations to determine whether or not this behavior is characteristic of $O(n)$, $n\ge 2$ models, taking into account the fact that the coefficient in front of the $\ln L$ term is quite small in the spherical model---see Eq. \eqref{eq:bform}.

The qualitative agreement between our study and the experiments on  $^4$He encourages us to anticipate that the model we investigated can prove to be a very useful adjunct to general, and perhaps specific,  studies of Casimir forces in systems with a continuous symmetry of the type that is broken in the superfluid transition. Note that in \cite{GC99} some spreading is reported in the scaled plots of the measured Casimir force acting on helium films of different thickness formed on 
Cu plates. Later in \cite{GSGC2006} where $^4$He films formed on a silicon surface have been studied this spreading is greatly reduced, and the previously observed effect on Cu has been attributed solely to the roughness of the Cu surfaces used in \cite{GC99}. One might speculate that an additional reason for the spreading is the  existence of $\ln L$ corrections to the scaling behavior of the force.  

We have shown that when $T\to 0$ the scaling function of the Casimir force tends to a universal constant; see Eqs. \eqref{F0} and \eqref{d1}. This implies that the Casimir force tends to zero in that limit, as the force is linear in $T$. Of course, our purely classical approach has ignored quantum fluctuations. Given the existence of zero point motion, a properly quantized system with gapless modes, should manifest a non zero Casimir force at $T=0$. One promising candidate for the investigation of the Casimir force at low temperatures when the governing fluctuations are of quantum rather than thermal origin are the different versions of the quantum spherical model \cite{O72,N95,V96,CDT2000,VZ92}. The finite size behavior of one version of this model has been studied  under periodic boundary conditions \cite{CDT2000}. We hope that our analytical results will make it possible to study this, and related to it models, such as the quantum anharmonic crystal \cite{VZ92}, subject to Dirichlet boundary conditions.

\appendix
\section{On the properties of matrix $\mathcal{M}$}
\label{sec:matrix_M}
First, since the elements of this matrix are real and $\mathcal{M}_{ij}=\mathcal{M}_{ji}$ one knows that the eigenvalues, $a_l$, are real, the eigenvectors are orthonormal $\braket{\psi^{(l)}}{\psi^{(m)}}=\delta_{l,m}$, and that those eigenvectors form a complete system, i.e. \begin{equation}
\label{completeness}
\sum_{l=1}^L\left[\ket{\psi^{(l)}}\bra{\psi^{(l)}}\right]_{ij}=\sum_{l=1}^L\psi^{(l)}_i\psi^{(l)}_j=\delta_{ij}.
\end{equation}
Next, according to Eq. (\ref{DefinitionOfMMatrix}), one has
\begin{eqnarray}
\label{ADefinitionOfMMatrix}
\mathcal{M}_{ij}&=& 2\Lambda_i\; \delta_{i,j}-\delta_{|i-j|,1}\\
& = & 2\Lambda_{\rm max}\;\delta_{i,j}-\tilde{\mathcal{M}}_{ij}. \nonumber
\end{eqnarray}
where $\Lambda_{\rm max}=\max_{i}\Lambda_i$ and 
\begin{equation}
\tilde{\mathcal{M}}_{ij}=\left[2\left(\Lambda_{\rm max}-\Lambda_i\right)\; \delta_{i,j}+\delta_{|i-j|,1}\right]\ge 0.
\end{equation}
We now make use of the Perron-Frobenius theorem \cite{Se73,T2005,W78} concerning
the eigenvalues and eigenvectors of an indecomposable matrix $\textbf{A}$ of nonnegative elements $a_{i,j}\ge 0$. We express this property by writing $\textbf{A}\ge 0$. A matrix $\textbf{A}=\{a_{i,j}\}$ is called connected or indecomposable if for any two indices $i$ and $j$ there is a sequence $r_k$, $1\le k \le s$, such that the product $a_{i,r_1}a_{r_1,r_2}a_{r_2,r_3} \cdots a_{r_s,j}\ne 0$. If $\textbf{A} \ge 0$ is a real connected matrix, it has a largest simple positive eigenvalue,
$r(\textbf{A})=r$, and an associated column vector
$\textbf{x}>0$, such that $\textbf{A} \textbf{x}=r \textbf{x}$ where $r>0$; any
other eigenvalue $\lambda$ of $\textbf{A}$ has absolute value
less than or equal to $r$. Further, if $\textbf{B} \ge 0$ is
another real matrix of the same dimension,
such that $\textbf{A}-\textbf{B} \ge 0$, then $r(\textbf{B}) \le r(\textbf{A})$,
the equality holding only if $\textbf{B}=\textbf{A}$. Applying the above theorem to the matrix $\tilde{\mathcal{M}}$ we find that the matrix $\mathcal{M}$ has a non-degenerate smallest eigenvalue $a_1$, the corresponding  eigenvector having components that are all positive, i.e., $\psi_i^{(1)}>0$, $i=1,\cdots,L$. Furthermore, the following theorem \cite{F86,Be2009} holds: If $\textbf{A} = a_{i,j}$ is a real tridiagonal matrix of order $L$ satisfying $a_{k, k + 1} a_{k + 1,k} > 0$ for $k = 1,\cdots, L - 1$, then $\textbf{A}$ has $L$ real simple eigenvalues. Taking into account that for the elements of the matrix $\mathcal{M}$ one has $m_{k,k+1}=m_{k+1,k}=-1$ for $k = 1,\cdots, L - 1$, and,  therefore $m_{k,k+1}m_{k+1,k}=1>0$, we conclude that \emph{all} eigenvalues $a_l, l=1,\cdots, L$ of $\mathcal{M}$ are real and non-degenerate. From the general theory of tridiagonal matrices one can also gain some knowledge for the behavior of the eigenvectors associated with the corresponding eigenvalues. The following theorem \cite{F86} is valid:  Under the conditions of the previous theorem if $\lambda_1>\lambda_2\cdots>\lambda_L$ are the eigenvalues of $\textbf{A}$, which exist in virtue of the previous theorem, then every (real) eigenvector ${\mathbf z}=(z_1,z_2,\cdots,z_L)^{\rm T}$ of the matrix $\textbf{A}$ has the properties: {\it i)} $z_1\ne 0, z_L\ne 0$; {\it ii)}  If $z_k=0$ then $a_{k-1,k}a_{k,k+1}z_{k-1}z_{k+1}<0$; {\it iii)} If we delete zeros from the sequence 
$$ z_1, a_{1,2}z_2, a_{1,2} a_{2,3}z_3, \cdots, a_{1,2}a_{2,3}\cdots a_{n-1,n}z_n$$ 
and if the vector ${\mathbf z}$ belongs to $\lambda_r$, then there are exactly $r - 1$ changes of sign in the sequence. Applying this theorem to the matrix $\mathcal{M}$, taking into account that $m_{k,k+1}=m_{k+1,k}=-1$ for $k = 1,\cdots, L - 1$, we obtain, again, that the smallest eigenvalue $a_1$ is characterized by eigenvector ${\mathbf \Psi}^{(1)}$ with components $\psi_i^{(1)}>0$, $i=1,\cdots,L$, i.e., it has no zero elements. The next to the smallest $a_2$ eigenvalue corresponds to eigenvector ${\mathbf \Psi}^{(2)}$ the components of which  change sign once.  Due to  symmetry this occurs in the middle of the system. So, if we take a system with odd number of component $L$, the component with coordinate $i=(L+1)/2$ will be zero, i.e. $\psi_{(L+1)/2}^{(2)}=0$, while, say $\psi_i^{(2)}>0$ for $i=1,\cdots, (L-1)/2$ and $\psi_i^{(2)}<0$ for $i=(L+3)/2,\cdots, L$ . In the general case the eigenvector ${\mathbf \Psi}^{(r)}$ has $r-1$ changes of the sign of its subsequent components. As a result of the symmetry of the problem it is clear that the eigenvectors are either symmetric or anti-symmetric about to the middle of the system, i.e. that  $\psi_i^{(l)}=(-1)^{l+1}\psi_{L+1-i}^{(l)}$. Thus, if the equation (\ref{SphericalConstraintThreeDimensions}) is fulfilled for some $i = k$, then it is automatically fulfilled also for $i^{\prime} = L + 1 - k$.

In the limit $L\to \infty$ the system will be described by a matrix $\mathcal{M}_b$ with ${\Lambda_i}$ independent of the layer number $i$, i.e., with $\Lambda_i=\Lambda$. The $L\times L$ matrix $\mathcal{M}_L$ with diagonal elements $\Lambda_i=\Lambda$, as is well known, see, e.g. \cite{Be2009}, is characterized by eigenvalues $\hat{a}_l=\Lambda-2\cos[l\pi/(L+1)]$ and eigenvectors $\hat{\psi}^{(l)}_i=\sqrt{2/(L+1)}\sin[i\, l\pi/(L+1)]$, $i,l=1,\cdots, L$.

\section{Numerical Determination of Lagrange Multipliers $\Lambda_l$, $l=1,\cdots,L$}
\label{Numerics}

We aim to determine the Lagrange multipliers $\Lambda_l$, $l=1,\cdots, L$ such that the eigenvalues $a_l$,  and eigenvectors $\ket{\psi^{(l)}}$, $l=1,\cdots,L$, of the matrix ${\mathcal{M}}$ defined in Eq. \eqref{DefinitionOfMMatrix} satisfy Eqs. (\ref{SphericalConstraintThreeDimensions}).  Our approach is numerical.  The solutions will, obviously, depend both on the temperature and the size of the system, i.e. $a_l=a_l(T,L)$ and $\ket{\psi^{(l)}}=\ket{\psi^{(l)}(T,L)}$, $l=1,\cdots,L$. In order to solve Eqs. \eqref{SphericalConstraintThreeDimensions} near and above the critical temperature $T_c$ of the system,  we use the multidimensional Newton-Raphson method. For temperatures $x=(L/\xi_0^+)t \ll -1$ we apply a modification of this method taking into account that the lowest eigenvalue of the system $a_1$ approaches its limiting minimal allowed value of 4 exponentially rapidly in$L$ (see Eq. (\ref{a1})).  \\

\subsubsection{Multidimensional Newton-Raphson method}
\label{sec:MultiNRM}

We have to solve the equations 
\begin{equation}
\label{Newton}
{\mathbf f}({\mathbf\Lambda})=0
\end{equation}
where ${\mathbf f}=\{f_1,f_2,\cdots,f_L\}$ with 
\begin{equation}
\label{SM_Newton}
f_i({\mathbf\Lambda})=-R+\frac{2}{\pi}\sum_{l=1}^L \left[\psi^{(l)}_i\right]^2\frac{1}{a_l}K\left(\frac{4}{a_l}\right)
\end{equation}
for each $i=1,\ldots,L$. According to the Newton-Raphson method one starts with a suitable chosen set of $\Lambda_l$, $l=1,\cdots,L$,  ${\mathbf \Lambda}_{\rm old}$, and iteratively generates new values  ${\mathbf \Lambda}_{\rm new}$, where 
\begin{equation}
\label{NewtonGen}
{\mathbf \Lambda}_{\rm new}={\mathbf \Lambda}_{\rm old}-{\mathbf D}^{-1} \cdot{\mathbf f}({\mathbf\Lambda}_{\rm old}),
\end{equation} 
with ${\mathbf D}=\{D_{i,j}\}$, $i,j=1,\cdots,L$, where 
\begin{equation}
\label{derivative}
D_{i,j}=\partial f_i/\partial \Lambda_j.
\end{equation}
In order to implement the method, we must first compute derivatives of the constraint equations with respect to the spherical fields $\{\Lambda_l\}$. To accomplish that requires the derivatives of $a_l$ and $\ket{\psi^{(l)}}$ on ${\mathbf \Lambda}$. The {\it exact} results, familiar from first order perturbation theory---see, e.g., \cite{LLQM}---or the operator expansion---see, e.g., \cite{RBvol4}---are
\begin{equation}
\frac{\partial a_l}{\partial\Lambda_j}=2\left[\psi^{(l)}_j\right]^2 \ge 0,
\end{equation}
which tells us that $a_l$, $l=1,\cdots, L$, are nondecreasing functions of $\{\Lambda_l\}$, and
\begin{equation}
\frac{\partial\psi^{(l)}_i}{\partial\Lambda_j}=2\sum_{m\ne l}\frac{\psi^{(m)}_j\psi^{(l)}_j}{a_l-a_m}\psi^{(m)}_i.
\end{equation}
With the help of the above results one can compute the Jacobian determinant $\mathbf D$ for Newton's method to be
\begin{equation}
\label{Dmatrixelem}
D_{i,j}=
\frac{4}{\pi}\sum_{l=1}^L\sum_{m=1}^L\psi^{(l)}_i\psi^{(m)}_i\psi^{(l)}_j\psi^{(m)}_j d_{l,m},
\end{equation}
i.e., $D_{i,j}=D_{j,i}$, and 
\begin{equation}
\label{delem}
d_{l,m}=\frac{E(4/a_l)}{16-a_l^2}\delta_{m,l}+\frac{2K(4/a_l)}{a_l(a_l-a_m)}(1-\delta_{m,l}),
\end{equation}
with $E(k)$ being the complete elliptic integral of the second kind with modulus $k$. Due to the properties of eigenvalues $a_l$, $l=1,\cdots,L$ of matrix ${\mathcal M}$ presented in Appendix \ref{sec:matrix_M}, one has $a_l\ne a_m$ if $l\ne m$ and, therefore, $d_{l,m}$ are finite and well defined when $a_1>4$ with $a_1=\min_l \{a_l\}$, $l=1,\cdots,L$. The condition $a_l>4$ is physically necessary, because $\mathcal{H}(\mathbf{q})$ must be positive definite, see Eq. \eqref{HOuterProductExpansion}, for the free energy, Eq. (\ref{FiniteSizeFreeEnergy}), to exist.

Newton's method works very well at high temperatures, where the eigenvalues $a_l$ are comfortably larger than $4$. We see empirically that the smallest eigenvalue, $a_1$, gets arbitrarily close to $4$ as we approach low temperatures.  While it is not \begin{em}a priori\end{em} obvious that the system will be driven to $a_1\gtrsim 4$, this behavior can be anticipated from the known behavior of the corresponding bulk system at its critical point.  Mathematically, it is straightforward to understand why this occurs. When $T$ becomes small so that $R$ becomes large, the constraint equations, Eq. (\ref{SphericalConstraintThreeDimensions}), begin to rely on the divergence of $K(x)$, forcing an eigenvalue to approach $4$ from above. In fact, as it is shown in Appendix \ref{sec:low_T}, $a_1$ gets exponentially close to $4$---again, see Eq. (\ref{a1}). Newton's method is, unsurprisingly, unstable in this region because iterations of the procedure  often send the system into the unphysical regions with an eigenvalue below $4$.\\

\subsubsection{Modified Newton-Raphson method} 
\label{sec:ModNRM}

Once $a_1$ is close enough to $4$ that Newton's method  exhibits numerical instability, the problem can be solved to an excellent approximation by implementing the following changes:
\begin{itemize}
\item Replace $K(a_1/4)$ in the constraint equation with a new free coefficient, $C$.
\item Enforce the condition that $a_1=4$.
\end{itemize}
Specifically, the new constraints are \cite{note1}
\begin{equation}
0=f_i(\Lambda,C)=-R+C\left.\psi^{(1)}_i\right.^2+\frac{2}{\pi}\sum_{l=2}^L \left.\psi^{(l)}_i\right.^2\frac{1}{a_l}K\left(\frac{4}{a_l}\right)
\end{equation}
for $i=1,\ldots,L$, and an additional constraint
\begin{equation}
0=g(\Lambda,C)=a_1-4.
\end{equation}
These $(L+1)$ equations are to be solved for the $(L+1)$ variables $\{\Lambda_l\}$ and $C$. The $(L+1)\times(L+1)$ Jacobian is computed in the same way as before, but with the $(L+1)$-st column given by $\partial f_i/\partial C$ and the $(L+1)$-st row given by $\partial g/\partial\Lambda_j$.

Once we have the means to compute the $\{\Lambda_l\}$ for a given system size and temperature, we would like to construct the Casimir force making use of \eqref{ourexpression}. This involves taking a (discrete) derivative of free energy with respect to system size, and subtracting off the corresponding bulk free energy in order to capture the purely finite-size contribution. The details needed for the bulk model are given in Appendix \ref{sec:bulk_model}.

\section{Some properties of the bulk model}
\label{sec:bulk_model}

The properties of the bulk spherical model are investigated in detail in \cite{J72,BDT2000}. Here we summarize the results needed for the current study. We start with the expression for the bulk free energy density $f_b$, which reads
\begin{equation}
\label{BulkSphericalFreeEnergyDensity}
\beta f_{b}=\left\{\begin{array}{cc}
-R\Lambda_b+[\ln\left(R/\pi\right)]/2+I\left(\Lambda_b\right),& R\ge R_c\\
-3R+[\ln\left(R/\pi\right)]/2+I\left(3\right),& R\le R_c
\end{array}\right.
\end{equation}
where 
\begin{equation}
I=\frac{1}{16\pi^3}\int_{-\pi}^\pi dq_x\,dq_y\,dq_z\,\ln\left(\Lambda-\cos q_x-\cos q_y-\cos q_z\right),
\end{equation}
and for $R<R_c$ the parameter $\Lambda_b$ is to be determined from the bulk spherical field equation.
\begin{eqnarray}
\label{bulksphericalconstraint}
 R &=& W(\Lambda_b)\equiv \dfrac{1}{16\pi^3}\int_{-\pi}^\pi\frac{dq_x\,dq_y\,dq_z}{\Lambda_b-\cos q_x-\cos q_y-\cos q_z}\nonumber \\
 &=& \dfrac{1}{2}\int_{0}^{\infty} dw\; e^{-w\Lambda_b}I_0^3(w)\\
 &=&\dfrac{1}{2\pi ^2} \int_0^{\pi } \frac{4}{ 2\Lambda_b -2\cos (q)} K\left(\frac{4}{2\Lambda_b -2\cos
     (q)}\right)\, dq. \nonumber
\end{eqnarray}
Here $R_c\equiv W(3)$ is given in Eq. \eqref{bulk_constraint} and  $I_0(w)$ is the modified Bessel function of the first kind. The last line in Eq. \eqref{bulksphericalconstraint} provides a representation that alludes the analogy with the finite-dimensional system. 

The behavior of the integral $I$ for $\Lambda=3$ was studied by Joyce and Zucker\cite{JZ2001}, and they succeeded in computing it to 51 digits,
\begin{equation}
I_3 \equiv I(\Lambda=3) \approx 0.4901210612051\ldots\quad.
\end{equation}
We note that
\begin{equation}
\frac{dI}{d\Lambda}=W(\Lambda)
\end{equation}
is the well-studied ``generalized Watson integral''. Fisher and Barber developed a series expansion of this integral \cite{BF73} for $\Lambda\approx 3$, showing that
\begin{equation}
W(\Lambda)=R_c-\frac{1}{4\pi}\sqrt{2(\Lambda-3)}+O(\Lambda-3).
\end{equation}
 Integrating with respect to $\Lambda$, we find the series expansion for $I(\Lambda)$,
\begin{equation}
I(\Lambda)= I_3+R_c(\Lambda-3)-\frac{\sqrt{2}}{6\pi}(\Lambda-3)^{3/2}+O\left((\Lambda-3)^2\right),
\end{equation}
valid when $\Lambda \gtrsim 3$. This series can be used in the region in which numerical evaluation of the integral $I(\Lambda)$ is slow and inaccurate.

\section{On the derivation of the Casimir force in the near under-critical and in the Goldstone mode dominated regime}
\label{sec:low_T}

We aim to solve Eqs. \eqref{SphericalConstraintThreeDimensions}  in the  regime $4\pi (R-R_c)\gg \ln L/L$. This relationship holds when the absolute temperature,  $T$, is a fixed distance below $T_c$ and $L$ is sufficiently large. In addition, it holds when $R-R_c$ vanishes as $L \rightarrow \infty$ as long as the difference is asymptotically large compared to $L^{-1}$, the extent of the finite scaling regime, in that it is sizable compared to the width of the scaling regime multiplied by $\ln L$. Our goal is to determine the behavior of the Casimir force in this ``low temperature" regime corresponding to a range of temperatures in which the Goldstone modes provide the leading contributions to the force \cite{D2013,[{See also }] Bergknoff2012}. As we will see, these contributions again lead to $L^{-3}$ scaling of the Casimir force. In contrast, when Goldstone modes are absent and when the boundary conditions do not give rise to an interface  within the system, the Casimir force well below $T_c$ decays exponentially in $L$, as in case of the Ising model. 

Using the completeness of the eigenvectors $\ket{\psi^{(l)}}$, $l=1,\cdots,L$ and performing the sum of Eqs. \eqref{SphericalConstraintThreeDimensions} with respect to $i$, we arrive at
\begin{equation}
\label{SummingSphericalConstraintEquations}
R=\frac{1}{2\pi}\dfrac{1}{L}\sum_{l=1}^L\frac{4}{a_l}K\left(\frac{4}{a_l}\right).
\end{equation}
Comparison with Eq. \eqref{bulksphericalconstraint} yields the result that, when $L\to\infty$, one has 
\begin{equation}
\label{replacement}
\dfrac{1}{L}\sum_{l=1}^L \rightarrow \dfrac{1}{\pi}\int_{0}^{\pi}dq, \qquad \mbox{and} \qquad a_l \rightarrow 2\Lambda_b -2\cos (q).
\end{equation}
In the bulk limit the critical coupling $R_c$ is determined by setting the spherical field to its lowest allowed value, at which it remains for all $R\ge R_c$. We note that $(4/a)K(4/a)$ is a monotonically decreasing function of the parameter $a$ that tends to $+\infty$ when $a \to 4^+$.  This tells us that, as $L$ increases, the lowest eigenvalue $a_1$ will approach the value $4$ from above as $R$ increases above  $R_c$. 

Let us assume that $a_1\to 4$ and determine, by self-consistency, the region in which that happens.  As already noted in the main text, in this regime $\mathcal{M}\to \mathcal{M}^*$, which is constructed according to (\ref{DefinitionOfMMatrix}), with the $\Lambda_i$'s replaced by the $\Lambda^*_i$'s in (\ref{lamst}),  the eigenvalues of $\mathcal{M}^*$ being given by Eq. \eqref{eq:aslams}---and the eigenvectors by Eq. \eqref{eq:pert4}. We now turn to the behavior of  $\mathcal{M}$ when its diagonal elements are close to $\mathbf{\Lambda}^*$. We consider perturbation of $\mathbf{\Lambda}^*$ of the form $\mathbf{\Lambda}^* \longrightarrow\mathbf{\Lambda}^*+\vec{\varepsilon}/2$ where $\vec{\varepsilon}=\{\varepsilon_1,\varepsilon_2,\cdots,\varepsilon_L\}$. Then it is straightforward to show that 
\begin{eqnarray}
\label{SphericalAsymptoticPerturbedEigenvalues}
a_l &=& \lambda_l+\sum_{i=1}^L\varepsilon_i\left[\phi^{(l)}_i\right]^2 \nonumber \\
&&+\sum_{i=1}^L\sum_{j=1}^L \varepsilon_i\varepsilon_j\sum_{k\ne l}\dfrac{\phi^{(l)}_i\phi^{(l)}_j\phi^{(k)}_i\phi^{(k)}_j}{\lambda_l-\lambda_k},\nonumber  \\ && + O(\varepsilon^2) 
\end{eqnarray}  
and 
\begin{eqnarray}
\label{SphericalAsymptoticPerturbedEigenvectors}
\ket{\psi^{(l)}}&=&\ket{\phi^{(l)}}+\sum_{m\ne l}\frac{\sum_{j=1}^L\varepsilon_j\phi^{(m)}_j\phi^{(l)}_j}{\lambda_l-\lambda_m}\ket{\phi^{(m)}} \nonumber \\ && +O(\varepsilon^2).
\end{eqnarray}
where $\lambda_l$ and $\phi^{(l)}$ are defined in (\ref{eq:aslams}) and (\ref{eq:pert4}). 

{\textit{\textbf{Derivation of the behavior of $a_1$}}

Let us start by determining the behavior of $a_1$ which we will accomplish without relying on  perturbation theory. It is necessary to proceed in this way because the function $g(x)$ which enters the equations is singular when $a_1\to 4^+$, i.e., it does not possess a Taylor-type expansion around the corresponding non-perturbative value of 4. In order to determine $a_1$ we study the behavior of Eq. \eqref{SummingSphericalConstraintEquationsshort} under the assumption that $a_1\to 4^+$ and that $g(a_1)\gg g(\lambda_2)$. Then, taking into account the fact that the term with $a_1$ provides the leading contribution to the sum we have 
\begin{equation}
R = \frac{1}{2 \pi L }K\left(\frac{4}{a_1} \right) + \rho_L, \label{eq:pert9}
\end{equation}
where 
\begin{multline}
\label{rhoL}
\rho_L\equiv\frac{1}{L}\sum_{l=2}^L  g\left( \lambda_l\right)\\=\frac{1}{2\pi L}\sum_{l=2}^L\frac{1}{1+\sin^2\left(\frac{\pi(l-1)}{2L}\right)}K\left(\frac{1}{1+\sin^2\left(\frac{\pi(l-1)}{2L}\right)}\right)\\
\to\frac{1}{2\pi}\int_0^1\frac{dx}{1+\sin^2\left(\frac{\pi x}{2}\right)}K\left(\frac{1}{1+\sin^2\left(\frac{\pi x}{2}\right)}\right)=R_c,
\end{multline}
i.e., $\rho_L\to R_c$ when $L\to\infty$. Inserting this result for $\rho_L$ in Eq. \eqref{eq:pert9} and expanding $K$ for $a_1\to 4^+$, one obtains the result reported in Eq. \eqref{a1} in the main text.  Eq. \eqref{a1} also exhibits the fact that $a_1\to 4$ when $4\pi L(R-R_c)\gg 1$. The result for $\rho_L$ can be further improved. Applying the improved Euler-Maclaurin formula  \cite{CG2008,S2004,Si2012} for functions with a logarithmic singularity at one end of the interval to the sum in Eq. \eqref{rhoL} (see especially Theorem 5 in \cite{CG2008}), one can show that 
\begin{equation}
\label{rolLimproved}
\rho_L=R_c-\frac{1}{4\pi L}\left(\frac{K(1/2)+7\ln 2}{2}+\ln L\right)+O\left(\frac{1}{L^2}\right).
\end{equation}
The condition $g(a_1)\gg g(\lambda_2)$, which we have imposed in the derivation of the behavior of $a_1$ leads, in turn, to the constraint 
\begin{equation}
\label{g1g2a}
4\pi (R-R_c)L\gg \ln L.
\end{equation}
Since from Eq. \eqref{g1g2}  $a_1\to4^+$  is also satisfied, Eq. \eqref{g1g2a} represents the main constraint for the validity of Eq. \eqref{a1_final}.

{\textit{\textbf{Derivation of the behavior of $a_l$, $l=2,\cdots,L$}}

We now turn to the task if obtaining the behavior of the eigenvalues $a_l$ for $l=2,\cdots,L$. To that end we will use Eqs. \eqref{SphericalConstraintThreeDimensions}. Supposing again $g(a_1)\gg g(\lambda_{2})$, $g(\lambda_{2})=\max_l g(\lambda_{l})$ for $l=2,\cdots,L$ (here we use the fact that $g(x)$ is a monotonically decreasing function of $x$), one obtains 
\begin{equation}
R = \left[\psi_i^{(1)}\right]^2 g(a_1) + C_i \label{eq:pert5}
\end{equation}
where 
\begin{equation}
C_i= \sum_{l=2}^L \phi_i^{(l) \, 2} g(\lambda_l). \label{eq:pert7}
\end{equation}
Obviously $C_i$, $i=2,\cdots,L$ are easily computed functions of only $L$. Our tactical goal is, using the orthonormality and the completeness of the eigenvectors $\ket{\phi^{(l)}}$, ${l=1,\cdots,L}$,  to determine $\varepsilon_i$, $i=1,\cdots,L$,  after inserting \eqref{SphericalAsymptoticPerturbedEigenvectors} in \eqref{eq:pert5} and keeping in the resulting equation only up to linear terms in $\varepsilon_i$, $i=1,\cdots,L$. Since this is a standard operation in perturbation theory, we simply report the final result: 
\begin{eqnarray}
\label{es}
\varepsilon_m &=& \dfrac{2}{L(R-\rho_L)}\sum_{l=2}^{L}g(\lambda_l)\\ & &\times \sin^2[\dfrac{\pi(l-1)}{L}]\cos[\dfrac{\pi(l-1)(2m-1)}{L}]. \nonumber
\end{eqnarray}
Using then, for $l=2,\cdots,L$, up to first order in $\varepsilon_i$, $i=1,\cdots,L$ Eq. \eqref{SphericalAsymptoticPerturbedEigenvalues} , one derives the expression given in Eq. \eqref{al_final} for the eigenvalues $a_l$. It is easy to check that 
\begin{equation}
\label{sum_e}
\sum_{m=1}^L \varepsilon_m=0. 
\end{equation} 
This, together with Eq. \eqref{SphericalAsymptoticPerturbedEigenvalues}
 demonstrates that, formally, within perturbation theory, one would simply have $a_1=\lambda_1$, while the nonperturbative solution yields $a_1$ given by Eq. \eqref{a1_final}. 

{\textit{\textbf{Derivation of the behavior of the Casimir force}}

In order to derive an analytical expression for the Casimir force we will use Eq. \eqref{FiniteSizeFreeEnergy} reported in the main text.
We will take there  
\begin{equation}
\label{some_def}
a_l=\lambda_l+\Delta_l, \qquad \mbox{and} \qquad 2\Lambda_l=2-\delta_{1,l}-\delta_{L,l}+\varepsilon_l,
\end{equation}
where, according to Eqs. \eqref{a1_final} and \eqref{al_final}
\begin{equation}
\label{delta1}
\Delta_1=32\exp[-4\pi(R-\rho_L)],
\end{equation}
\begin{equation}
\label{deltal}
\Delta_l=\frac{\sin^2[\pi(l-1)/L]}{L(R-\rho_L)}
\left[g(\lambda_l)-g(L+2-l)\right],
\end{equation}
for $l=2,\cdots,L$ and $\varepsilon_l$, $l=1,\cdots,L$ are given by Eq. \eqref{es}. Since we have derived $\Delta_1$ with precision of the order of $[L(R-\rho_L)]^{-1}$, it is this precision with which we are going to determine the Casimir force.  
Let us first deal with the sum 
\begin{equation}
\label{S}
S=\frac{1}{2}\sum_{l=1}^L\Bigg[\ln a_l
-\frac{2}{a_l^2}\cdot{}_4F_3\left(1,1,\frac{3}{2},\frac{3}{2};\,2,2,2;\,\frac{16}{a_l^2}\right)\Bigg]
\end{equation}
in Eq. \eqref{FiniteSizeFreeEnergy}. 
We start by noting two integral identities which will turn out to be helpful. First, it is easy to check that the generalized hypergeometric function ${}_4F_3$ in Eqs. \eqref{FiniteSizeFreeEnergy} and \eqref{S}
is related to the following  \cite{AS} logarithmic integral via
\begin{eqnarray}
\lefteqn{\label{log_integral}
\frac{1}{(2\pi)^2}\int_{-\pi}^\pi dx\,\int_{-\pi}^\pi dy\,\ln\left(s-2\cos x-2\cos y\right)}\nonumber \\
&&=\ln s - \frac{2}{s^2}\cdot{}_4F_3\left(1,1,\frac{3}{2},\frac{3}{2};\ 2,2,2;\ \frac{16}{s^2}\right).
\end{eqnarray}
Performing the derivative with respect to $s$ from the both sides of the above equation, or doing the calculations directly, one also obtains the following result for the Watson type two-dimensional integral \cite{BF73,JZ2001}
\begin{equation}
\label{W2d}
\frac{1}{(2\pi)^2}\int_{-\pi}^\pi dx\,\int_{-\pi}^\pi dy\,\frac{1}{\left(s-2\cos x-2\cos y\right)}=g(x),
\end{equation}
where $g(x)$ is given by Eq. \eqref{gf_def}. Then we find that $S$ can be approximated as 
\begin{equation}
\label{sumS}
S = S_0+S_1+S_{1,1},
\end{equation}
where 
\begin{equation}
\label{S0}
S_0=\sum_{l=1}^{L}\frac{1}{2(2\pi)^2}\int_{-\pi}^\pi dx\,\int_{-\pi}^\pi dy\,\ln\left(\lambda_l-2\cos x-2\cos y\right)
\end{equation}
will be responsible for the zero-temperature $L$-behavior of the force, while 
\begin{equation}
\label{S1}
S_1=\sum_{l=2}^{L}\frac{1}{2(2\pi)^2}\int_{-\pi}^\pi dx\,\int_{-\pi}^\pi dy\,\dfrac{\Delta_l}{\lambda_l-2\cos x-2\cos y}
\end{equation}
and 
\begin{eqnarray}
\label{S11}
\lefteqn{S_{1,1}=}\\
& &\frac{1}{2}\frac{1}{(2\pi)^2}\int_{-\pi}^\pi dx\,\int_{-\pi}^\pi dy\,\ln\left(\dfrac{a_1-2\cos x-2\cos y}{\lambda_1-2\cos x-2\cos y}\right) \nonumber,
\end{eqnarray}
will yield portions of its $T$-dependence. One can immediately deal with $S_{1,1}$. One finds that 
\begin{equation}
\label{s11der}
S_{1,1}=\frac{1}{2}\int_{\lambda_1}^{a_1} g(x) dx. 
\end{equation}
Taking into account that $g(s)$ is, in fact, the two-dimensional Watson type integral $W_2(s-4)$ and using its property  \cite{BF73,JZ2001} for $(s-4)\to 0^+$ that $W_2(s-4)\simeq \ln(s-4)/(4\pi)+5\ln2/(4\pi)+O(s)$, from Eqs. \eqref{delta1} and \eqref{s11der} it immediately follows that 
\begin{equation}
\label{s11_final}
S_{1,1}\simeq 16L(R-\rho_l)\exp[-4\pi L(R-\rho_L)]. 
\end{equation}
We are not going to determine the Casimir force with such an exponential precision, so, we will neglect the contribution to it stemming from $S_{1,1}$. 

{\textit{Derivation of the size dependence of $S_0$}

The $L$-dependence of $S_0$ can be determined exactly. To that end we make use of the identity, see Eq. 1.396.1 in \cite{GR}:
\begin{equation}
\label{ID_GR}
\prod_{k=1}^{n-1}\left(x^2-2x\cos\frac{\pi k}{n}+1\right)=\dfrac{x^{2n}-1}{x^2-1},
\end{equation}
which, with the substitution $x=\exp(v)$, can be written in the form
\begin{equation}
\label{ID_GR_final}
2^n\prod_{k=0}^{n-1}\left(\cosh(v)-\cos\frac{\pi k}{n}\right)=2\sinh (nv)\tanh(\frac{v}{2}).
\end{equation}
Taking into account the explicit form of $\lambda_l$, $l=1,\cdots,L$ given by Eqs.  \eqref{eq:aslams} and identifying $v$ from Eq. \eqref{v}, we derive from Eq. \eqref{S0}
\begin{equation}
\label{S0final}
S_0=\frac{1}{2(2\pi)^2}\int_{-\pi}^\pi dx\,\int_{-\pi}^\pi dy\, \ln\left[2\tanh\left(\frac{v}{2}\right)\sinh(Lv)\right].
\end{equation}
Thus, for the total pressure between the surfaces of the system due to the $S_0$ contribution into the free energy one has 
\begin{equation}
\label{F0total}
\beta F_{\rm tot}^{(0)}(L)=-\frac{\partial S_0}{\partial L}=-\frac{1}{2(2\pi)^2}\int_{-\pi}^\pi dx\,\int_{-\pi}^\pi dy\; v\coth(Lv).
\end{equation}
From Eq. \eqref{F0total} one  derives the corresponding result for the part of the Casimir force reported in Eq. \eqref{F0}.

{\textit{Derivation of the size dependence of $S_1$}

The sum $S_1$ can be written in the form 
\begin{equation}
\label{SphericalAsymptoticsB1asRiemannSum}
S_1=\frac{1}{(R-\rho_L)}\cdot\frac{1}{L}\sum_{m=1}^{L-1}G\left(\frac{m}{L}\right),
\end{equation}
where
\begin{multline}
G(x)=\frac{1}{8\pi^2}\frac{\sin^2(\pi x)}{1+\sin^2(\pi x/2)}K\left(\frac{1}{1+\sin^2(\pi x/2)}\right)\\
\times\left[\frac{1}{1+\sin^2(\pi x/2)}K\left(\frac{1}{1+\sin^2(\pi x/2)}\right)\right.\\
\left.-\frac{1}{1+\cos^2(\pi x/2)}K\left(\frac{1}{1+\cos^2(\pi x/2)}\right)\right].
\end{multline}
It is easy to check that $G(x)$ has logarithmic type singularities both near $x=0$, as well as near $x=1$. Therefore, in order to find the $L$-dependence of the sum $S_1$ one needs a modification of the standard Euler-Maclaurin summation formula, valid when the function
of interest has logarithmic singularities at its endpoints. Such a generalization of the Euler-Maclaurin summation formula has been recently proposed in \cite{CG2008} and \cite{S2004,Si2012} (see, e.g., theorem 2.1. in \cite{Si2012}). Applying the corresponding theorem one directly obtains 
\begin{eqnarray}
\lefteqn{S_1=\frac{1}{(R-\rho_L)}\left\{\int_{0}^{1}G(x)dx+\frac{1}{8L^3}
\Bigg[\zeta''(-2)\right.}\\
&&+\left(2\ln L-K\left(\frac{1}{2}\right)+7\ln 2-2\ln(2\pi)\right)\zeta'(-2)\Bigg]\nonumber\\
&& +\frac{\pi^2}{192L^5}\Bigg[-14\zeta''(-4)+\left(5-4E\left(\frac{1}{2}\right)+11K\left(\frac{1}{2}\right)\right.\nonumber\\
&&\left.\left.-98\ln 2+28\ln(2\pi)-28\ln L\right)\zeta'(-4)\Bigg]
\right.+{\cal O}(L^{-7})\Bigg\}. \nonumber 
\end{eqnarray}
Then for the corresponding contribution of $S_1$ towards the Casimir force in which we will retain only terms of the order of $(R-\rho_l)^{-1}$ one obtains {\it two} times the result reported in Eq. \eqref{d2}. As we will see, half of the  $L$-dependence of $S_1$ is also contained in the $R$-proportional term in the free energy given by Eq. \eqref{FiniteSizeFreeEnergy}. Let us now deal with this term. One has
\begin{equation}
\label{Rterm}
\sum_{l=1}^{L}\Lambda_l=2L-2+\sum_{l=1}^{L}\varepsilon_l,
\end{equation}
where we have used Eq. \eqref{some_def}. According to Eq. \eqref{sum_e} the last sum over $\varepsilon$'s is zero and thus, it looks like that this term does not contribute to the Casimir force up to the order of $[L(R-\rho_L)]^{-1}$, which we have retained in our previous calculations. However, the sum over $\varepsilon$'s is multiplied by $R$ and we require an expression for  $\sum_{l=1}^{L}\varepsilon_l$ up to the order $[L(R-\rho_L)]^{-2}$ in order to determine whether this sum contributes to the behavior of the Casimir force calculated up to the order of $[L(R-\rho_L)]^{-1}$. We now briefly describe how one can derive the  perturbation result needed. One starts again from Eqs. \eqref{eq:pert5} and \eqref{eq:pert7} but uses there the corresponding perturbation expansion for $\ket{\psi^{(l)}}$ up to second order in $\varepsilon$'s. Then one considers a small perturbation $\eta_l$ to any $\varepsilon_l$, as given by Eqs. \eqref{es}. Next, one uses the standard procedures within  perturbation theory and after some tedious, cumbersome, but otherwise straightforward calculations obtains that the $L$-dependent part of the sum $\sum_{l=1}^{L}\eta_l$ is half of that of $S_1$. The overall conclusion then is, that up to the order of $[L(R-\rho_L)]^{-1}$ the Casimir force is as reported in Eq. \eqref{d2}.


\begin{thebibliography}{112}
\bibitem [{not({\natexlab{a}})}]{note2}%
  \BibitemOpen
  \href@noop {} {}\bibinfo {howpublished} {The GLW approach involves expansion
  about small values of the amplitude of the order parameter field. This is a
  step in the standard mapping procedure of any microscopic model onto the GLW
  functional. For low T the amplitude of the order parameter is not small and,
  therefore, the approximation made will produce large deviations from the
  behavior of the realistic microscopic model} \BibitemShut
  {NoStop}%
\bibitem [{not({\natexlab{b}})}]{note3}%
  \BibitemOpen
  \href@noop {} {}\bibinfo {howpublished} {H. E. Stanley demonstrated
  \cite{S68} that the free energy density of the classical $n$-vector model in
  the thermodynamic limit converges precisely to that of the spherical model
  when the spin dimensionality $n$ tends to infinity. In addition, all the
  correlation functions of the $n$-vector spin model converge as $n \to \infty$
  to the corresponding correlation functions of the spherical model \cite{S88}.
  For a detailed review of the subsequent efforts in this direction, say for
  systems with disorder, we refer the reader to \cite{KKPS92}}
 \BibitemShut {NoStop}%
\bibitem [{not({\natexlab{c}})}]{note4}%
  \BibitemOpen
  \href@noop {} {}\bibinfo {howpublished} {The film considered here does not
  undergo an ordering transition at any finite temperature which means that,
  strictly speaking, symmetry is not broken at non-zero temperature. However,
  we do find that the gap in the lowest lying modes of excitation is extremely
  small at low temperatures (see the results in Sec.\ref{Section_Asymptotics},
  especially the behavior of the quantity $a_1-4$, which is effectively the
  above-mentioned gap). This means that the notion of Goldstone modes, the
  gapless branch of the excitation spectrum that follows from a breaking of
  continuous symmetry, applies quantitatively in a low temperature regime}
  \BibitemShut {NoStop}%
\bibitem [{not({\natexlab{d}})}]{note1}%
  \BibitemOpen
  \href@noop {} {}\bibinfo {howpublished} {Note that the procedure described
  here is mandated when the $d-1$ dimensional system undergoes an ordering
  system, as when $d \geq 4$. Then the critical point of the finite system
  $T_{c,L}$ is marked by sticking of $a_1$ to its minimal allowed value 4. One
  has $a_1=4$ for $T \le T_{c,L}$} \BibitemShut {NoStop}%
\end{thebibliography}

%

\end{document}